\documentclass{aa501}

\usepackage{epsf,epsfig}

\begin{document}

%
\title{Rapid variability of accretion  in AM Herculis}


\author{ D. de Martino\inst{1},
G. Matt\inst{2},
B.T. G\"ansicke\inst{3},
R. Silvotti\inst{1},
J.M. Bonnet-Bidaud\inst{4},
M. Mouchet\inst{5,6}
}

\offprints{D.~de Martino}

\institute{
Osservatorio Astronomico di Capodimonte, Via Moiariello 16, I-80131 Napoli, 
Italy
\email{demartino@na.astro.it, silvotti@na.astro.it}
\and
Dipartimento di Fisica, Universita' degli Studi Roma Tre, Via della
Vasca Navale, ROMA, Italy
\email{matt@fis.uniroma3.it}
\and
Astronomy Group, University of Southampton, Hampshire, UK
\email{btg@astro.soton.ac.uk}
\and 
Service  d'Astrophysique Saclay Gif-Sur-Yvette, France
\email{bobi@discovery.saclay.cea.fr}
\and
LUTH FRE2462 du CNRS, Observatoire de Paris, Section de Meudon,
F-92195 Meudon Cedex, France
\and
Universit\'e Denis Diderot, Place Jussieu F-75005, Paris, France
\email{martine.mouchet@obspm.fr}
}

\date{Received 25 July 2002; Accepted 9 September 2002}

\authorrunning{de Martino et al.}
\titlerunning{Rapid variability of accretion in AM Herculis}
\markboth{...}{...}

\abstract{ 
We present the last pointed observation of AM\,Her carried out during  
the life of the 
BeppoSAX satellite. It was bright at the beginning of the
observation, but dropped to the lowest X-ray level ever
observed so far. The X-ray emission during the bright period is consistent
with accretion occurring onto the main pole of the magnetized white dwarf.
The rapid change from the {\em active state} to the low deep state
indicates a drop by a factor of 17 in the accretion rate and hence that 
accretion switched-off. The short timescale (less than one hour) of this
variation still remains a puzzle. Optical photometry acquired
simultaneousy
during the low state shows that the white dwarf remains heated, although
a weak emission from the accretion stream could  be still
present. Cyclotron radiation, usually dominating the V and R bands, is
negligible thus corroborating the possibility that AM\,Her was in an
off-accretion state. The X-ray emission during the inactive state is
consistent with coronal emission from the secondary late type star.
}

\maketitle 

\section{Introduction}

AM Her is the prototype of Polars, strongly magnetic Cataclysmic
Variables (mCVs) (10-230\,MG), and consists of a magnetized
($\sim$ 14\,MG) white dwarf 
accreting from a late type (M4V) Roche lobe-filling secondary star.
Polars are characterized by long-term (months to years) high and low
accretion states which, due to
the absence of an accretion disc in these systems, reflect changes in the
mass loss rate of the donor star. AM Her is the brightest and
best  monitored Polar in the optical range and hence represents a test
object to study the evolution of the instantanous mass accretion
rate with time and then to understand the causes of the mass transfer
variations from the secondary star.  Bright and
faint luminosity states ($\Delta V \sim$ 2-3\,mag) occur on irregular
timescales from less than a day to months.  Different models have been
discussed to account for the long-term  mass transfer variations in CVs
(Livio \& Pringle 1994; King \& Cannizzo 1998) and the occurrence of
starspots at the inner Lagrangian point appears to be the most
likely
explanation. Along this line, Hessman et al. (2000) derived the mass
transfer rate history of AM Her, using the long term optical variability
and convert it into starspot filling factors. They conclude that the
density of the starspots  near the L1 point is unusually high (about
50$\%$).

On the other hand, monitoring of the X-ray activity  has been
relatively sparce, but has
already brought further insights into the variability of the mass
accretion rate. In particular, since its launch, we undertook a programme
with the BeppoSAX satellite to monitor the  X-ray behaviour of AM Her, to
infer  the evolution of the X-ray luminosity and its X-ray spectral
variability. In this work we present new and the most recent observations
of AM Her
carried out during the life of BeppoSAX together with simultaneous optical
photometry
acquired at the Loiano Bologna Observatory, which reveal further new
results on the
accretion variability and on the possible identification of the coronal
X-ray emission from the magnetically active secondary star.

\section{Observations and data reduction}

During its life the BeppoSAX satellite (Boella et al. 1997) 
performed  with the co-aligned Narrow Field Instruments (NFI) 
four pointed   observations of  AM\,Her. The source was found
to be in  different states as summarized in Table\,1 and depicted in 
Fig.\,1,  where the secular  AAVSO optical light curve
is shown along with the times of the X-ray pointings.

\begin{table*}[t]     
\centering 
\caption{History of BeppoSAX observations of AM Her and log of optical
observations in 2001.}
\vspace{0.05in}
\begin{tabular}{lccc}
\hline
\hline
   ~ &          ~ &      ~  &   ~ \cr
Date & MECS Exp. Time & Flux$^{1}$/Count Rate$^{2}$ & Notes \cr
     &   (s)          &                             & \cr
\noalign {\hrule}
   ~ &              ~ &                              ~ & \cr
Sep. 6, 1997 & 24700$^{3}$ & 0.18/0.049 & Active state {\em (a)} \cr
Sep. 6, 1997 &       & 0.024/6.5$\times10^{-3}$ & Quiescent low state {\em
(a)} \cr
May  8, 1998 & 33500 & 1.80/0.22   & Intermediate state {\em (b)} \cr
Aug. 12, 1998 & 80600 & 12/1.35    & High state {\em (b)} \cr
Apr. 22, 2001 & 30500$^{3}$ & 1.05/0.135 & Active state {\em (c)} \cr
Apr. 22, 2001 &       & 0.0154/4.5$\times10^{-3}$ & Quiescent low state {\em (c)} \cr
    ~ &             ~ &          ~ &    ~ \cr
\noalign {\hrule}
   ~ &              ~ &                              ~ & \cr
Date &  Filter       & Exp. Time  & Tintegration \cr
   ~ &              ~ &  (s) & (s) \cr
\noalign {\hrule}

   ~ &              ~ &                              ~ & \cr
Apr. 22, 2001 & U     & 11300 & 15 \cr
              & B     & 11300 & 5 \cr
              & V     & 11250 & 10 \cr
              & R     & 11250 & 20 \cr
   ~ &              ~ &                              ~ & \cr
\hline
\hline
\end{tabular}
~\par
\begin{flushleft}
$^{1}$: 2-10\,keV phase averaged flux in units of
10$^{-11}$\,erg\,cm$^{-2}$\,s$^{-1}$, see references for best fit
models.\par
$^{2}$: Count rates in units of cts\,s$^{-1}$.\par
$^{3}$: Total on source exposure time.\par
{\em a}: de Martino et al. (1998).\par
{\em b}: Matt et al. (2000) \par
{\em c}: This work. 
\end{flushleft}
\end{table*}

\begin{figure}[h]
\mbox{\epsfxsize=9cm\epsfbox{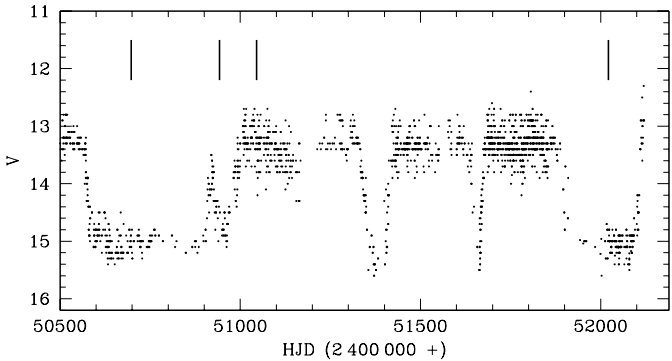}}
\caption[]{\label{curves} The long term optical light curve from 
AAVSO data. The times of BeppoSAX NFI pointings are  marked
with vertical lines.}
\end{figure}

\subsection{The BeppoSAX data}

The last BeppoSAX pointing of AM\,Her was carried out on April 22, 2001
during a low state that started in December 2000 and terminated at the end 
of July 2001 (Fig.\,1). It was detected at its lowest level by  
the Low Energy Concentrator Spctrometer (LECS) [0.1-10\,keV] and by the 
Medium Energy Concentrator Spectrometer (MECS) [1.3-10\,keV] with an
effective on source exposure time of 15.5\,ks and 30.5\,ks respectively.

Spectra and light curves from the MECS and LECS instruments
have been extracted from a
 circular region with a radius of 3' and 2' respectively, using the Ftools
XSELECT procedure. Background data
were extracted from blank sky pointings  using the same radius and
subtracted from  the data. 
  The BeppoSAX observation initially shows AM\,Her at a higher count rate
level
(henceforth {\em ''active state''}) which was monitored for $\sim$
2.2\,hr, followed by a deep low 
count rate state monitored for $\sim$ 16.1\,hr (henceforth {\em
''quiescent state''}).  In  Fig\,2, the LECS and MECS light curves show a
decrease by a factor of 47 and 30 in count rates between the  {\em active}
and {\em quiescent states} respectively.

\begin{figure}
\mbox{\epsfxsize=9cm\epsfbox{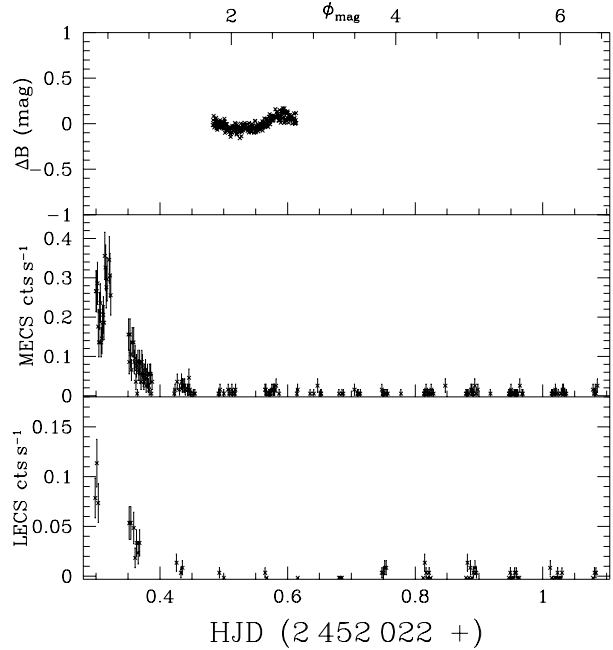}}
\caption[]{\label{curves} The light curves from the LECS (lower panel),
the MECS (middle panel) instruments.  Gaps are due  to the BeppoSAX 
satellite orbit. The B band differential photometry acquired
simultaneously is shown in the upper panel. Magnetic phases are also 
reported on the top. 
}
\end{figure}

\subsection{The optical data}

Optical UBVR photometry was carried out at the 1.5\,m telescope of
Loiano-Bologna Observatory (Italy) with a three channel photometer
(TTCP) on the
same day of BeppoSAX observation, covering simultaneously the {\em
quiescent state } for $\sim$
3.14\,hr (Fig.\,2, upper panel). Data reduction was performed using
the  TTCP software developed at the Astronomical Observatory of
Capodimonte-Naples. A set of Landolt standard stars were observed to
calibrate data resulting in an average U=14.43$\pm$0.04\,mag, B=15.43$\pm$
0.1 and V=14.99\,mag $\pm$ 0.04\,mag. R band data were not calibrated 
due to unadequate detector efficiency estimates and  poor calibration
data. 
The optical level during these observation is similar to that observed
during the low state in September 1997 (V=15.08\,mag).

\section{Data analysis and results}

The observations were analyzed separately  
according to the BeppoSAX count rate level. 

\subsection{The X-ray active state}

During the {\em active state} AM\,Her shows a first peak or flare 
followed by a second one and finally a fast exponential decay ($\propto
e^{-t/\tau}$, with $\tau$ = 51\,min, in the MECS detector). Due to the
lower efficiency, the LECS coverage is poorer. 
A similar fast decay was recorded by BeppoSAX during the low state in
September 1997 (de Martino et al. 1998), but at that time the {\em active
state} count rate level was three times lower than the one observed in
April 2001. 

\noindent We have folded the MECS {\em
active state} 2001 data along the 3.1\,hr rotational/orbital period using
the magnetic ephemeris reported in Heise \& Verbunt (1988), and compared
with the high state light curve observed by BeppoSAX in 1998 (Matt et
al. 2000). Although fragmentary, the light curve shown in Fig.\,3 shows
that the two peaks find their counterparts in the high state
light curve. They are
consistent with the bright phase (maximum) and hence with the X-ray
emission from the main accreting pole which is self-occulted at
$\phi_{mag}$=0.1-0.3. However the decay occurs somewhat earlier (by
$\sim$ 0.1 in phase), possibly
suggesting the concurrence of the beginning of disappearance of the main
accreting pole and of an intrinsic drop in the accretion flux.  
It is worth noting that the X-ray {\em active state} observed during
September 1997 was also compatible with X-ray emission onto the main
accreting pole (de Martino et al. 1998).\\

\begin{figure}
\mbox{\epsfxsize=9cm\epsfbox{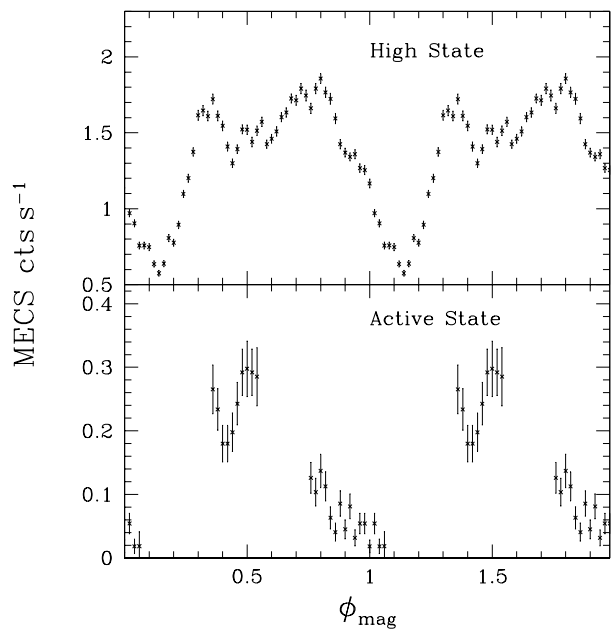}}
\caption[]{\label{curves} The MECS folded light curve of the active state
(lower panel) and that of the high state in 1998 (Matt et al. 2000) (upper
panel) shown for comparison (see text).}
\end{figure}

The averaged combined LECS and MECS spectrum during the {\em
active state}  does not require a black-body component, si\-mi\-lar to
that of September 1997. The spectrum was then fitted using the XSPEC
package with a MEKAL model  with iron abundances fixed at the solar value  
(Matt et al. 2000)  plus absorption fixed at $\rm N_{H}= 9\times
10^{19}\,cm^{-2}$
(G\"ansicke et al. 1995), which gives $\rm kT=14^{+23}_{-6}$\,keV
($\chi^{2}_{red}$=1.38 for 44 d.o.f). 
Though the fit is poor (Fig.\,4, left panel), the optically thin plasma
temperature is compatible with that found during intermediate and high
states (Matt et al. 2000). The data are not of enough quality to allow a
partial covering
component to be added. The bolometric flux during the {\em active state}
is $\rm 2.6\times 10^{-11} erg\,cm^{-2}\,s^{-1}$. 

\begin{figure*}
\epsfig{file=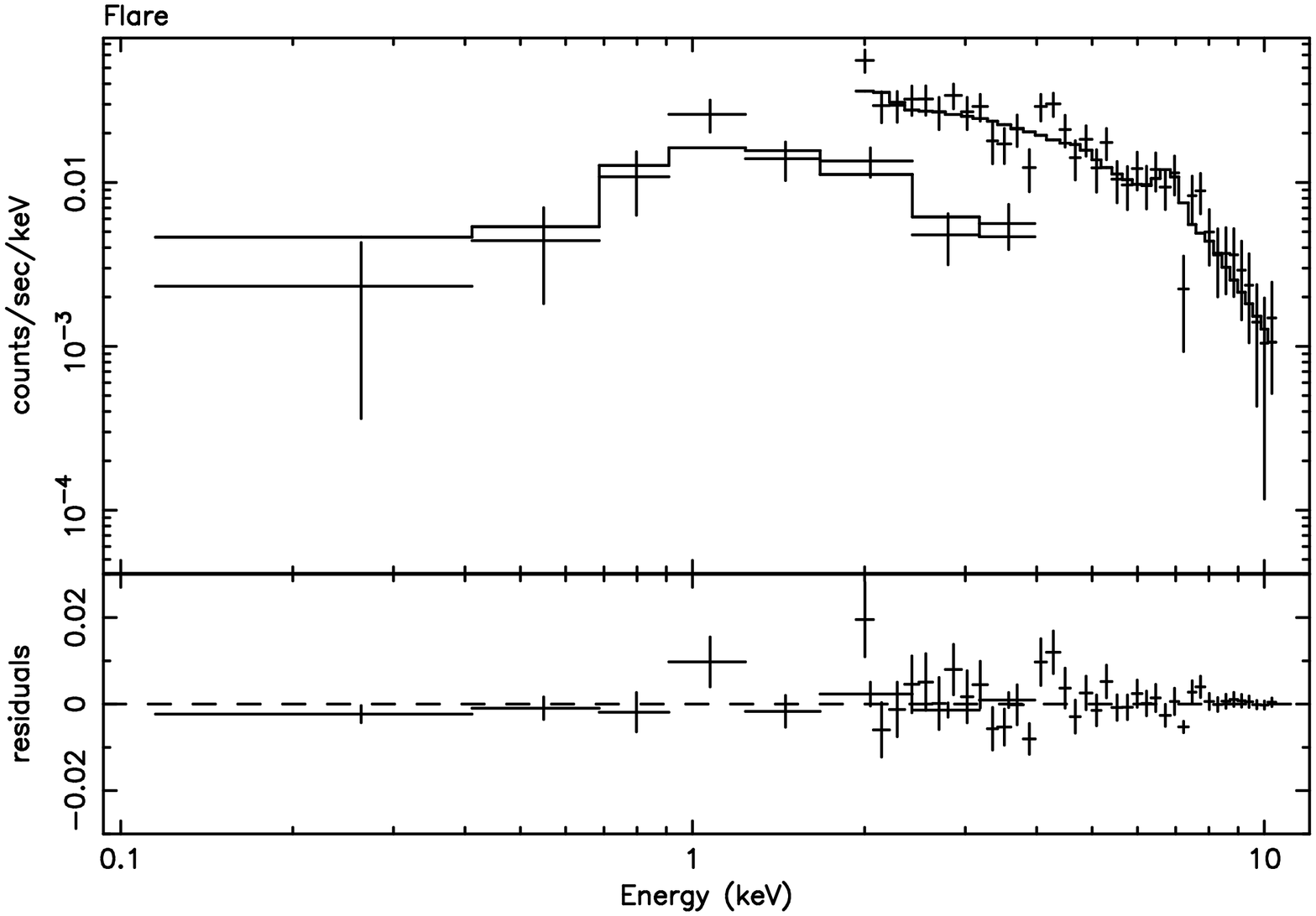, height=6.cm, width=8.5cm,angle=-90}
\epsfig{file=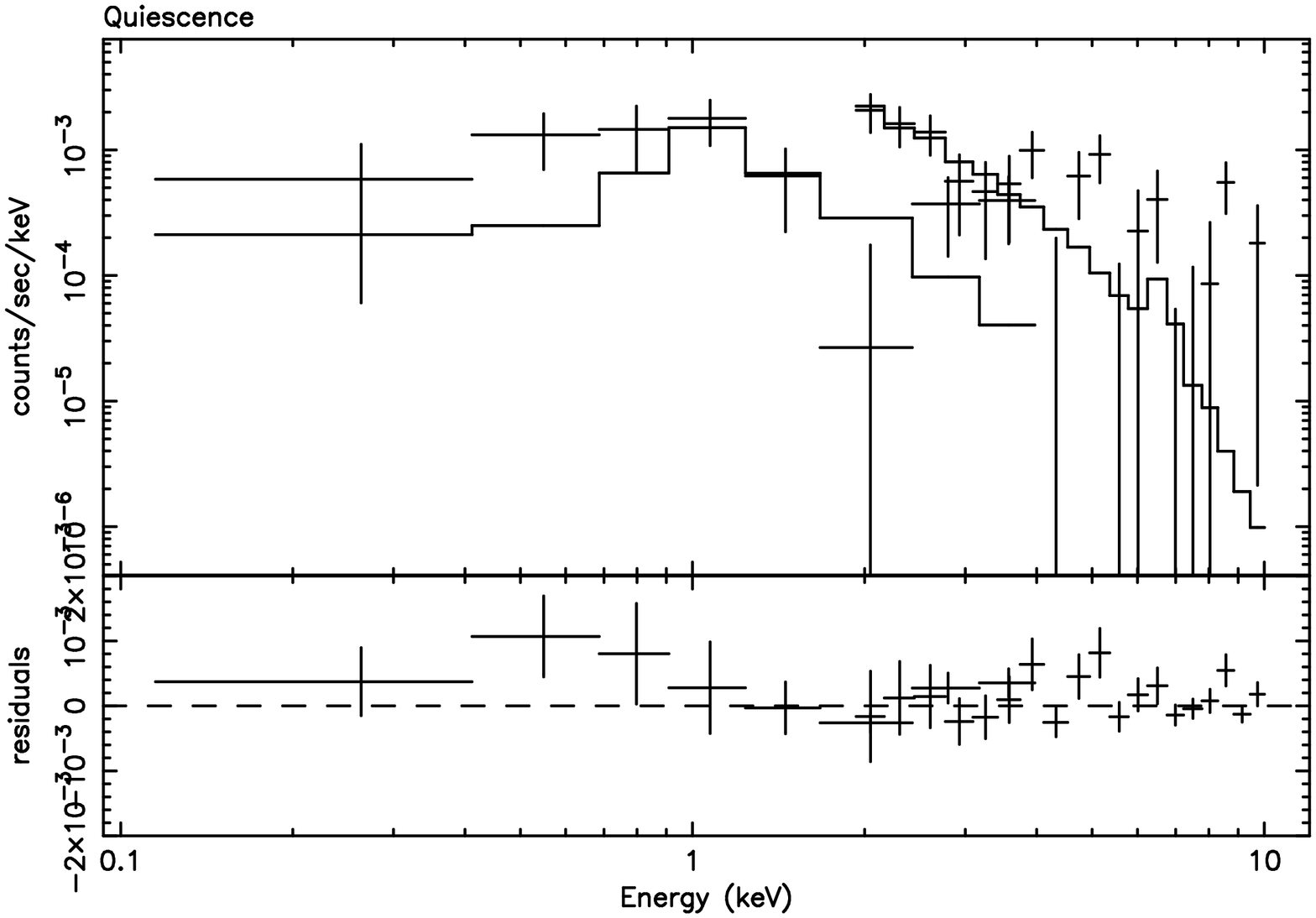, height=6.cm, width=8.5cm, 
angle=-90}
\caption[]{\label{curves} The combined LECS and MECS spectra during 
the {\em active state} {\it (left panel)}  and the {\em quiescent
state} {\it (right panel)} fitted  with MEKAL models (see
text).}
\end{figure*}

\subsection{The quiescent state}

\subsubsection{The X-ray emission}

While during the {\em active state}, the  count rate is larger than
that observed in the September 1997 {\em active state}, this is not the
case
for the  {\em quiescent state } count rate, which, in
September 1997, was $\sim$ 1.4 times higher than the one recorded in
2001, thus making the present BeppoSAX data the faintest X-ray state
detected so far in AM\,Her.  

\noindent No orbital modulation can be detected in the MECS data, thus
suggesting
that accretion onto the main pole has switched off. \\

The {\em quiescent state} average combined LECS and MECS spectrum
(Fig.\,4, right panel) was
 fitted with the same MEKAL plus absorption model which gives a much
lower temperature kT=1.5$\pm ^{+0.9}_{-0.7}$\,keV ($\chi^{2}_{red}$=1.43,
for
24 d.o.f), the lo\-west
determined so far in AM\,Her.  The bolometric
flux is $\rm 1.5\times 10^{-12}\,erg\,cm^{-2}\,s^{-1}$. 

\begin{figure}
\mbox{\epsfxsize=9cm\epsfbox{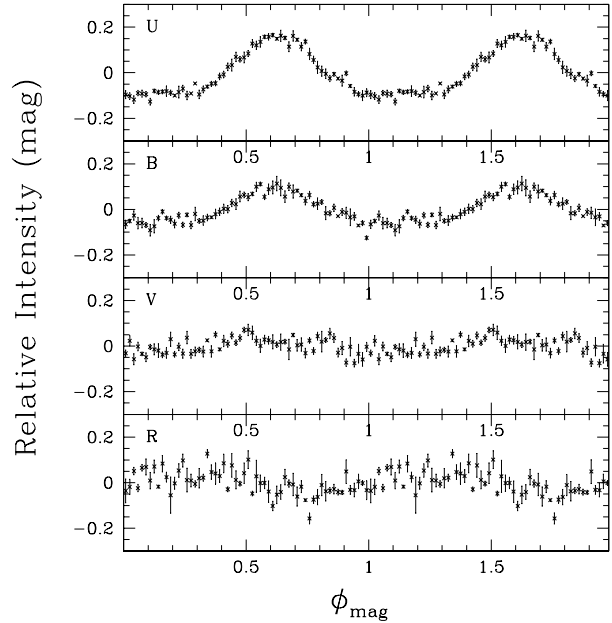}}
\caption[]{\label{curves} The differential UBVR folded light curves
obtained simultaneously during the X-ray {\em quiescent state}.}
\end{figure}

\subsubsection{The optical quiescent light curves}

Unlike the X-rays, the optical light curve shows an
orbital modulation (Fig.\,2). Folding differential photometric UBVR data
along the orbital period, a clear colour dependence is found, as depicted
in Fig.\,5. The amplitude of modulation changes from 20$\%$ in the U band,
to 13$\%$ in the B and is almost absent in V and R bands. The blue filters
show the typical behaviour observed in the far-UV range (G\"ansicke et
al. 1995; G\"ansicke et al. 1998) with a maximum at $\phi_{mag}$=0.64 and
a minimum at $\phi_{mag}$=0.1. This is consistent with the emission from
the white dwarf plus the heated pole cap which dominates the UV range
during both high and low accretion states (G\"ansicke et al. 1995). The
lack of significant modulation in V and R bands  suggests that cyclotron
emission, known to be dominant at these wavelengths, is negligible during
this low state. 

Following the method described in  G\"ansicke et al. (2001), we have
attempted to model the calibrated UBV  light curves. The first
attempt assumes that the U band light curve is due entirely to emission
from the white dwarf plus the heated polar spot. We fix the following 
parameters to the values derived from the previous works (G\"ansicke et
al. 1995, 1998):  distance $d= 90$\,pc, the white dwarf radius 
$R_\mathrm{wd}=1\times 10^{9}$\,cm, inclination $i = 50^{\circ}$,
colatitude and azimuth of the polar cap $\beta=55^{\circ}$ and
$\Psi=0^{\circ}$.
Cyclotron emission is neglected as well as the emission
from the accretion stream. The secondary M4V star contribution is taken 
into account as described in G\"ansicke et al. (2001) and set to
U=19.76\,mag, B=19.43\,mag and V=16.83\,mag. 
The temperature of the white dwarf is then adjusted to match the
faint-phase flux
which agrees  with the UV IUE low state results (G\"ansicke et
al. 1995). The central spot temperature and the spot opening angle are
then adjusted to match the amplitude and width of the observed modulation
in the U band, thus giving $\rm T_{wd}$=19000\,K, $\rm
T_{spot}$=50000\,K,
$\Theta_\mathrm{spot}$ = 20$^{\circ}$. The model, however, overpredicts
the
observed
B band flux and underpredicts the V flux. 

\noindent A match with the observed light curves is found, as
shown in Fig.\,6, by lowering the temperature ($\rm T_{wd}$=17000\,K) of
the white dwarf in order to decrease its contribution. The
spot temperature and opening angle are the same as before since they 
represent the amplitude and shape of the modulation well. However,
a constant component (with $U$=16.1\,mag) has to be added. This could be
attributed to the optically thin emission of a weak and faint accretion
stream. Also, a very weak $V$=18.5\,mag component  is required to 
 match the V flux. This could be due to a faint contribution
from cyclotron emission still present. The lack of a clear modulation in
the R band light curve (Fig.\,5) is not easy to interpret, although the
data cannot exclude that a modulation at a 10$\%$ is present. 
In the red, ellipsoidal variations from the
secondary star of about 0.1-0.2\,mag are expected (Bailey et
al., 1988) together with the heated white dwarf  atmospheric spot
modulation.  A possibility  could
be due that the weak cyclotron emission, whose behaviour is in 
anti-phase with 
the white dwarf spot modulation (see G\"ansicke et
al. 2001 for modeling of cyclotron contribution, is present producing a
flat R band light curve. Unfortunately the lack 
of reliable calibration in this band does not allow us to properly assess
this issue.

\begin{figure}
\mbox{\epsfxsize=9cm\epsfbox{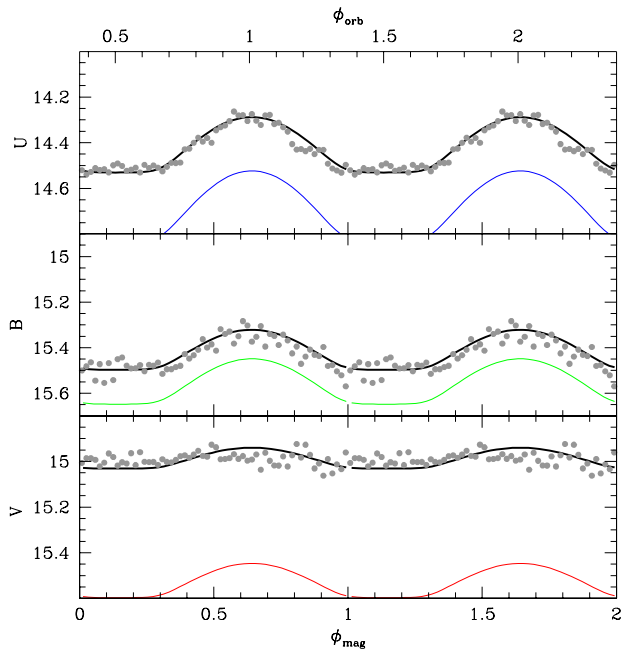}}
\caption[]{\label{curves} The observed  UBV light curves during the
 X-ray {\em quiescent state} 
together with the fitted model as described in the text. The solid thick
line
is the sum of all contributions from the heated white dwarf spot at
50000\,K, the un-heated white dwarf at 17000\,K plus a constant
contribution
from the accretion stream. In the V band an additional weak 18.5\,mag
component is included. The solid thin line represents the heated white
dwarf spot. The other contributions are not shown due to the large
magnitude differences. See text for details. }
\end{figure}

\section{Discussion and conclusions}

The last pointed BeppoSAX observation of AM\,Her caught the source
during its deepest low state ever observed. The X-ray flux shows a rapid
variability from an {\em active} to a {\em quiescent state} similar to
that observed during another prolonged low state in 1997
(de Martino et al. 1998). Unfortunately,
also for this second time, AM\,Her was bright at the beginning of the 
observation and, although the bright state is consistent with X-ray
accretion-induced emission, it is not possible to assess whether this was
a temporary accretion event or if the source was previously 
in a  constant accretion epoch. However, the drop in X-ray flux indicates
that the accretion rate decreased by a factor of $\sim$ 17 in less than
one hour. A very rapid variation (18\,min) in the X-ray flux also has been
 recorded in XMM-Newton data of UZ\,For during a deep low X-ray state 
and interpreted as an accretion event (Still \& Mukai
2001).  For the {\em active
state}, we derive an estimate of the accretion luminosity, assuming that
about half of the thermal bremsstrahlung and cyclotron radiation emitted
from the post-shock region is intercepted by the white dwarf and
re-emitted in the UV ($\rm L_{UV}=L_{tb} + L_{cyc}$) and neglecting the
contribution of a re-processed component in the EUV range
(cfr. G\"ansicke et al. 1995 and similar reasoning in de Martino et
al. 1998). We then estimate an accretion rate of $\rm 4.9\times
10^{-11}\,M_{\odot}\,yr^{-1}$.  The drop in mass
accretion rate therefore indicates that the X-ray emission during the {\em
quiescent
state} is not due to accretion. The much lower temperature
(1.5\,keV) is compatible with coronal temperatures of M type dwarfs
(Schmitt et al. 1990) as well as the 
low X-ray luminosity in the 0.05-3\,keV of $\rm 8.6\times
10^{29}\,erg\,s^{-1}$ is  compatible with those observed in active
late type stars (Pallavicini et al. 1990). Furthermore the emission
measure derived during the {\em quiescent state} EM= $\rm 4.6\times
10^{52}\,cm^{-3}$ is also consistent with coronal values. We have also
compared the X-ray luminosity during the {\em quiescent state} with
that expected from coronal emission of rapidly rotating late type
stars. For a secondary star filling its Roche lobe in a 3.09\,hr orbital
period binary, the period-radius relation (Patterson 1984) gives for the
secondary star
$\rm R_{sec}=0.32-0.36\,R_{\odot}$. The predicted saturation value for the
X-ray luminosity due to  rotation is $ \rm \propto R_{sec}^{2}$ (Fleming et
al. 1989), corresponding to $\rm 4.9-6.3\times
10^{29}\,erg\,s^{-1}$ (0.3-3\,keV band), in
 remarkable agreement with the luminosity derived in the same range
for the BeppoSAX {\em quiescent state} ($\rm 6.4\times
10^{29}\,erg\,s^{-1}$). 

\noindent Moreover, the lack of
V and R band modulations indicate that, even if present, cyclotron
emission is very weak, thus implying that AM\,Her really switched-off
accretion during the BeppoSAX observation. 
All this indicates that the X-ray
emission can be indeed and more safely ascribed to the secondary star
than
done previously for the {\em quiescent state} in September 1997. 

\noindent As for the 1997 BeppoSAX data set, still remains 
unclear  the cause of the rapid drop of accretion flux, with
timescales remarkably close to the dynamical timescale of the 
secondary star.  The 
observations of a  rapidly evolving burst in UZ\,For observed with
XMM-Newton further
confirms that such rapid changes occur in AM\,Her stars and a coronal mass
ejection event at the L1 point may be 
the cause of the observed active state  (cfr. 
de Martino et al. 1998 for discussion).  It
is clear that secondaries in these
close  binary systems are still far from being understood and further
X-ray observations during low states would help in assessing the nature of
these active secondary stars.

\begin{acknowledgements}
In this work, we have used and gratefully acknowledge the data from the
AAVSO International Database, based on observations submitted to the AAVSO
by variable star observers worldwide. 
DdM and GM acknowledge financial support from ASI. BTG was supported by a
PPARC advanced Fellowship.
\end{acknowledgements}


\end{document}